\documentclass[final]{appolb}
\usepackage{epsfig}
\usepackage{graphicx}
\def\runhead{An indication for deconfinement at RHIC}

\begin{document}
\title{AN INDICATION FOR DECONFINEMENT IN Au+Au COLLISIONS AT RHIC%
\thanks{Presented at XXXIIIth Int. Symp. Multipart. Dyn.
    Cracow, Poland, 9/5-11/2003}%
}
\author{M. Csan\'ad$^1$, T. Cs\"org\H{o}$^2$, B. L\"orstad$^3$
    and \underline{A. Ster}$^2$
\address{$^1$ Dept. Atomic Physics, ELTE, H-1117 Budapest, P\'azm\'any P. 1/a, Hungary \\
$^2$ MTA KFKI RMKI, H - 1525 Budapest 114, POBox 49, Hungary \\
$^3$ Dept. Physics, University of Lund, S - 22362 Lund, Sweden}
}
\maketitle
\begin{abstract}
We present simultaneous, quality fits to final BRAHMS, PHENIX,  PHOBOS and STAR
data on particle spectra and two-particle Bose-Einstein or
HBT correlations as measured in Au+Au collisions at $\sqrt{s_{NN}}
= 130$ GeV. Using the Buda-Lund hydro model, 
the best fit is achieved when the most central 1/8th 
of the particle emitting volume is superheated to  $T > T_c =170$ MeV.
In contrast, we find no indication for such a hot central region in a
similar analysis of Pb+Pb data at CERN SPS.
\end{abstract}

\section{Introduction}
Recently, Fodor and Katz calculated the phase diagram
of lattice QCD  at finite net barion density~\cite{Fodor:2001pe}.
The results, obtained with light quark masses about
four times heavier than the physical value,
indicated that in the $0 \le \mu_B \le 200$ MeV region the transition
from confined to deconfined matter is a cross-over,
with $T_c \simeq 172 \pm 3$ MeV which is, within errors, independent of the
bariochemical potential in this region.
At the same time, deconfined matter is searched for in Pb+Pb collisions
 with laboratory bombarding energies of 158 AGeV at CERN SPS as well as
in Au + Au collisions at RHIC, with $\sqrt{s_{NN}} = 200$ GeV
colliding energies.

\section{Basic properties of the Buda-Lund hydro model}
 Here we report on an analysis of the measured
single particle spectra and two-particle Bose-Einstein correlation functions
using the Buda-Lund hydrodynamical model~\cite{Csorgo:1995bi,Csorgo:1995vf}.
The key features of this model can be summarized as follows. 

1) The particle emitting source is assumed to have a core-halo structure~\cite{Csorgo:1994in},
where the particle production in the core is governed by local
thermalization and relativistic, three-dimensional expansion, while particle
production in the halo is governed by the decays of long-lived hadronic
resonances, that have lifetimes larger, or much larger
than the characteristic 5-10
fm/c lifetimes of the hydrodynamically evolving core.
In case of pion production, such resonances are e.g. $\omega$, $\eta$ and
$\eta^\prime$.

2) The freeze-out distribution of the core is described by a three
- dimensionally expanding, locally thermalized source, that
reflects the symmetry properties of the collision (axial symmetry
for central collisions).

3) The model goes back to known solutions of (relativistic) hydrodynamics
in limiting cases when such solutions are known (e.g. non-relativistic
or ultrarelativistic limits).

4) The flow field has a transverse and a longitudinal Hubble flow,
the density and the inverse temperature distributions are parameterized
keeping the means and variances of these distributions only, in such a form
that all the observables can be calculated analytically from the model.

5) If the flow vanishes  and the temperature  and density are constants,
the freeze-out temperature can be determined from the single-particle
spectra and the geometrical sizes of the source from two-particle
Bose-Einstein correlations. However, in the limit of a large, strongly
expanding source with local temperature inhomogeneities, the
two-particle correlations are dominated by thermal length scales
(sizes within which the Boltzmann factor is nearly constant),
and the single particle spectra contain (or even are dominated)
by flow and temperature inhomogeneity effects, corresponding to
a scaling limiting behaviour.

6) In the scaling limit, the analytic formulas given
in refs.~\cite{Csorgo:1995bi,Csorgo:1995vf,Csorgo:1999sj}
yield the analytic prediction that
$R_{out}\simeq R_{side} \simeq R_{long} \propto 1/\sqrt{m_t}$,
or, with other words,
$R_{out}/R_{side} \simeq R_{side}/R_{long} \simeq R_{long}/R_{out} \simeq 1$,
where the equality is reached for sufficiently large transverse masses,
satisfying $m_t \gg T_0$, and at lower $m_t$ the model predicts
scaling violating correction terms.

This  Buda-Lund hydro model and the above mentioned analytic formulas
were recently reviewed and summarized in ref.~\cite{Csorgo:1999sj},
as far as the single particle spectra and the two-particle Bose-Einstein
or HBT radius parameters are concerned.
Since at RHIC the BRAHMS and the PHOBOS collaborations reported  data
on the pseudo-rapidity distributions in refs.
~\cite{Bearden:2001xw,Back:2001bq},
 we have extended this model
and have improved the analytic calculation in such a way, that
in the longitudinal, space-time rapidity variable, the position
of the point of maximal emittivity can be found exactly now,
and thus analytic calculation is not anymore limited by the linearized
saddle-point equations.

\section{Fits to final run-1 Au+Au data at RHIC}
\begin{table}[b]
\begin{center}
\begin{tabular}{|l|rl|rl|rl|}
\hline
                 BL parameter
                 & \multicolumn{2}{c|}{Au+Au \@ RHIC}
                 & \multicolumn{2}{c|}{Pb+Pb \@ $\langle$SPS$\rangle$}
                 & \multicolumn{2}{c|}{ h+p  \@ SPS} \\ \hline
$T_0$ [MeV]           & \hspace{0.3cm} 214    &$\pm$ 7
                      & \hspace{0.3cm} 139    &$\pm$ 6
                      & \hspace{0.3cm} 140    &$\pm$ 3 \\
$\langle u_t \rangle$ & 1.0   &$\pm$  0.1
                      & 0.55   &$\pm$ 0.06
                      & 0.20   &$\pm$ 0.07 \\
$R_{s|G}$ [fm]        & 8.6    &$\pm$ 0.4
                      & 7.1    &$\pm$ 0.2
                      & 7.3    &$\pm$ 0.8  \\
$\tau_0$ [fm/c]       & 6.0    &$\pm$ 0.2
                      & 5.9    &$\pm$ 0.6
                      & 1.4    &$\pm$ 0.1  \\
$\Delta\tau$ [fm/c]   & 0.3    &$\pm$ 1.2
                      & 1.6    &$\pm$ 1.5
                      & 1.36   &$\pm$ 0.02  \\
$\Delta\eta$          & 2.3    &$\pm$ 0.4
                      & 2.1    &$\pm$ 0.4
                      & 1.36   &$\pm$ 0.02\\
$T_{\mbox{\rm s}}$ [MeV]
                  & 0.5 $T_0$   & fixed
                      & 131    &$\pm$ 8
                      &  82    &$\pm$ 7 \\
$T_{\mbox{\rm e}}$ [MeV]
              & 102    &$\pm$ 11
                      & 87     &$\pm$ 24
                      & 0.1    &$\pm$ 0.1 \\ \hline
\end{tabular}
\end{center}
\caption
{
Source parameters from simultaneous fits of
final BRAHMS, PHENIX, PHOBOS and STAR data for
$Au+Au$ collisions at $\sqrt{s_{NN}} = 130 $ GeV data,
as given in Fig. 1, obtained  with the Buda-Lund hydrodynamical model,
version 1.5. Pb+Pb data were fitted in ref.~\cite{blfit-CERN},
while h+p data at CERN SPS in ref.~\cite{Agababian:1997wd}.
}
\label{tab:results}
\end{table}

\begin{figure}[!thb]
\begin{center}
\vspace{-0.5cm}
  \begin{center}
\includegraphics[width=2.4in]{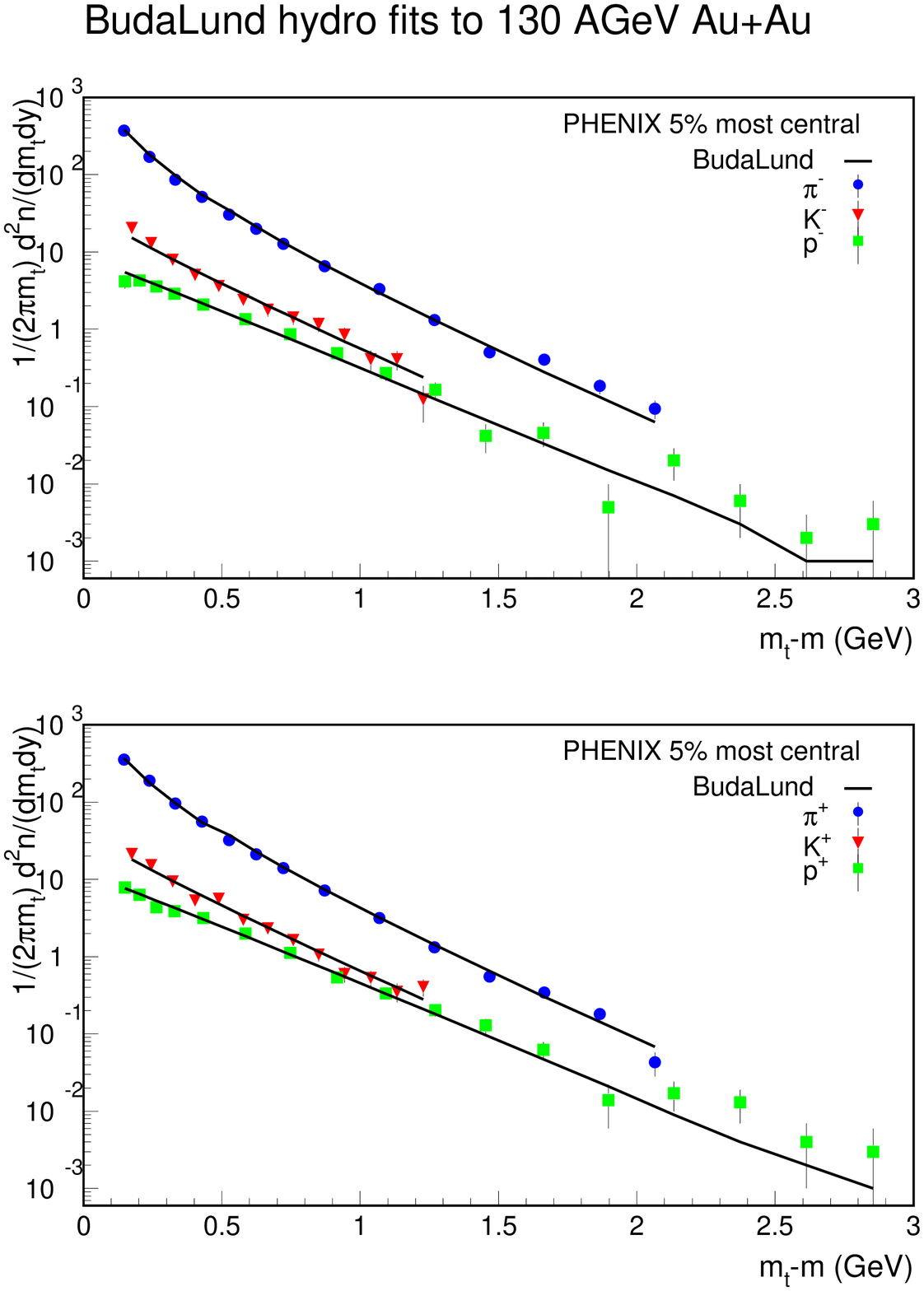}
\includegraphics[width=2.4in]{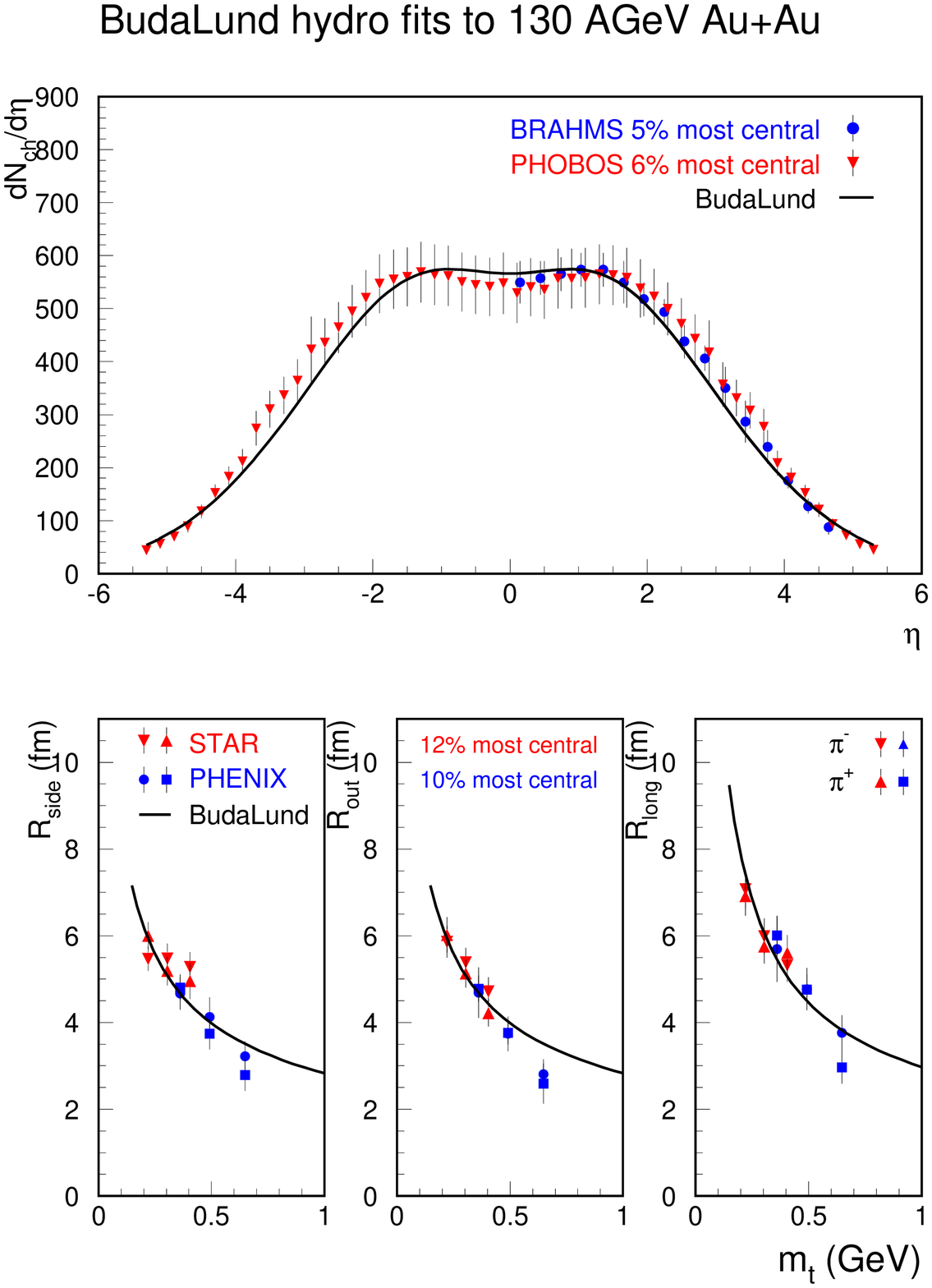}
\end{center}
\caption[*]{
\label{fig:spectra}{\small Solid line shows the simultaneous Buda-Lund fit  to final
Au+Au data at $\sqrt{s_{NN}} = 130$ GeV, vers. 1.5. The 
transverse mass distributions are measured by PHENIX~\cite{Adcox:2001mf},
the  pseudorapidity distributions of charged particles
are measured by BRAHMS~\cite{Bearden:2001xw} and PHOBOS~\cite{Back:2001bq},
while the transverse mass dependence of the radius parameters are
measured by STAR~\cite{Adler:2001zd}
and PHENIX~\cite{Adcox:2002uc}. } 

}
\end{center}
\end{figure}

\begin{figure}[!thb]
\begin{center}
\vspace{-0.5cm}
  \begin{center}
\includegraphics[width=3.2in]{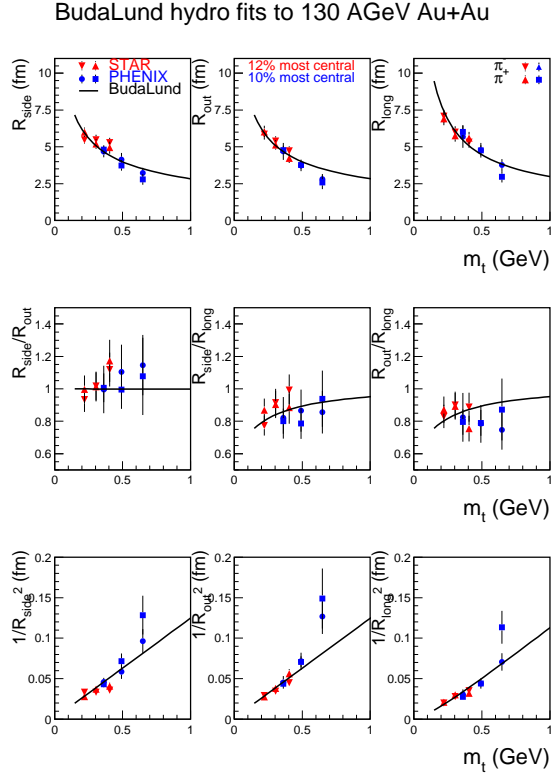}
\end{center}
\caption[*]{
\label{fig:radii} {\small Top row shows the transverse mass dependence of the
side, out and longitudinal HBT radii, the central line shows their
pairwise ratio (usually only $R_{\mbox{\rm out}}/R_{\mbox{\rm side}}$ is
shown) together with the  Buda-Lund fits, vers. 1.5. The bottom line shows
the inverse of the squared radii.
The intercept of the curves in this row is within errors zero for the
two transverse components, so the fugacity is independent of the
transverse coordinates. As the intercept is nonzero in the
longitudinal direction, the fugacity
(hence particle ratios) are rapidity dependent.}
}
\end{center}
\end{figure}
In this analysis, final BRAHMS~\cite{Bearden:2001xw},
PHENIX~\cite{Adcox:2001mf,Adcox:2002uc},
PHOBOS~\cite{Back:2001bq}, and STAR~\cite{Adler:2001zd} data are
used, and all these datapoints were fitted simultaneously, using
the analytic expressions and the CERN Minuit fitting package. 
The fitting package used in this analysis is version 1.5, 
made public at~\cite{Csorgo-blhome}.  A
well defined minimum was found, with accurate error matrix and a
statistically acceptable fit quality. The results are shown on
Fig. 1 and the essential parameters are summarized in Table 1. In
addition to the parameters in Table 1, we have 6 chemical
potentials that regulate the absolute normalization of the single
particle spectra of $\pi^\pm$, $K^\pm$, and $p^\pm$. The quality
of the fit to the RHIC data is characterized by $\chi^2$/NDF=
158/180, that corresponds to a confidence level of 88 \%. More
details are given in ref.~\cite{Ster-Warsaw}. Here we note only,
that the net baryochemical potential, $\mu_B$ can be determined
from the difference of the proton and antiproton chemical
potentials. In our case, this value is $\mu_B = 77 \pm 38$ MeV,
well within the range where the critical temperature is within
errors independent of the baryochemical potential according to
ref.~\cite{Fodor:2001pe}.
The fitting of RHIC data turned out to be more
complicated than that to the CERN SPS NA44, NA49 and WA98 data,
given in ref.~\cite{blfit-CERN}. The reason was that the RHIC HBT
correlation data are closer to the scaling limiting case than the
data at CERN SPS. This implied that the fugacity distribution,
$\exp(\mu(x)/T(x))$ becomes a constant parameter at RHIC, within
resolvable errors, independently of the transverse coordinate
values. Hence the ``geometric" contribution to the HBT radius
parameters is negligibly small. The finite transverse size seems
to be generated by the finite transverse size of the hot region,
but the fugacity term is within the resolvable errors independent
of the transverse coordinates in this region. This is signalled by
the fact that the fits became independent of the geometrical
source sizes, $R_G$, which characterizes the transverse variations
of the fugacity term, and we observed that $R_G$ becomes an
irrelevant parameter at RHIC, with $1/R_G^2 \approx 0$. This can be seen in the lower plots
of Fig. 2, where we show the inverse squared radii versus $m_t$.
The Buda-Lund model predicted, see eqs. (53-58) in 
ref.~\cite{Csorgo:1995bi} and also eqs. (26-28) in~\cite{Csorgo:2002kt}, 
that the linearity of the inverse radii as a function of $m_t$ 
can be connected to the Hubble flow and the temperature gradients. 
The slopes are the same for side,
out and longitudinal radii if the Hubble flow (and the temperature
inhomogeneities) become direction independent. The intercepts of the
linearly extrapolated $m_t$ dependent inverse squared radii at $m_t=0$ determine
$1/R_G^2$, or  the magnitude of corrections from the finite
geometrical source sizes, that stem from the $\exp(\mu(x)/T(x))$
terms. We can see on Fig. 3, that these corrections within errors
vanish at RHIC. This result is important, because it explains, why
thermal and statistical models can be successful at RHIC: if
 $\exp(\mu(x)/T(x)) = \exp(\mu_0/T_0)$, then this factor becomes
an overall normalization factor, proportional to the particle
abundances. Indeed, we found that when the finite size in the
transverse direction is generated by the $T(x)$ distribution,
the quality of the fit increased and we had no degenerate
parameters in the fit any more. In Table 1, we determined
$R_s$ at RHIC by the condition, that $T(r_x=r_y=R_s) = T_0/2$.

On the other hand we have obtained similarly good description of
these data if we require that the four-velocity field is a fully
developed, three-dimensional Hubble flow, with $u^\mu =
x^\mu/\tau$, however, we cannot elaborate on this point here due
to the space limitations~\cite{mate}.

\section{Conclusions}
Table 1 and Figures 1 and 2 indicate that there is no
real HBT puzzle at RHIC- the Buda-Lund hydro model works here as
well as at CERN SPS, and gives a good quality description of the
transverse mass dependence of the HBT radii. For the dynamical reason,
see refs.~\cite{Csorgo:2002kt} and \cite{Csorgo:1995bi}.  Naturally, we find
that the transverse size of the source is larger and the
transverse flow is stronger in heavy ion collisions than in h+p
collisions. We also observe, that the central temperature,
 $T_0 = 214 \pm 7$ MeV, is significantly higher at RHIC, than the critical
temperature, $T_c = 172 \pm 3$ MeV, calculated from lattice QCD,
and this is a significant, more than 5 standard deviation effect.
Finding similar parameters from the analysis of the pseudorapidity
dependence of the elliptic flow, it was estimated in
ref.~\cite{mate} that about 1/8th of the total volume is above the
critical temperature in Au+Au collisions at RHIC at the time when
pions are emitted from the source. We interpret this result as an
indication for quark deconfinement and a cross-over transition in
Au+Au collisions at $\sqrt{s_{NN}} = 130 $ GeV at RHIC. A similar
conclusion has been first achieved in ref.~\cite{Csorgo:2002ry}
when analysing the final PHENIX and STAR data on midrapidity
spectra and Bose-Einstein correlations, but only at a three
standard deviation level. By including the pseudorapidity
distributions of BRAHMS and PHOBOS, the significance of the result
increased.


\end{document}